\documentclass[aps,prl,amsmath,amssymb,superscriptaddress,twocolumn]{revtex4-1}

\ifx\pdfoutput\undefined
\usepackage{graphicx}
\else
\usepackage{graphicx}
\usepackage{bm}
\fi
\usepackage{wasysym}
\usepackage{xcolor}
\usepackage{color}

\usepackage{CJK}
\newcommand{\upd}{\mathrm{\,d}}

\begin{document}

\begin{CJK*}{GB}{gbsn} 
\title{Lagrangian single particle turbulent statistics through the  Hilbert-Huang Transform
}

\author{Yongxiang Huang (»ÆÓÀÏé)}%
\email{yongxianghuang@gmail.com}
\affiliation{ Shanghai Institute of Applied Mathematics and Mechanics Shanghai Key Laboratory of Mechanics in Energy Engineering Shanghai University, Shanghai 200072, People\rq{}s Republic of China}

\author{Luca Biferale}
\affiliation{ Department of Physics and INFN, University of Tor Vergata Via della Ricerca Scientifica 1 I-00133 Roma, Italy
}

\author{Enrico Calzavarini}
\affiliation{ Laboratoire de M\'ecanique de Lille CNRS/UMR 8107
Universit\'e Lille 1 59650 Villeneuve d\rq{}Ascq, France
}

\author{Chao Sun (Ëﳬ)} %
\affiliation{Physics of Fluids Group, University of Twente P.O. Box 217, 7500 AE Enschede, The Netherlands
}

\author{Federico Toschi}
\affiliation{Department of Physics, and Department of Mathematics and Computer Science and J.M. Burgerscentrum, Eindhoven University of Technology, 5600 MB Eindhoven, The Netherlands
}

\date{\today}

\begin{abstract}
The Hilbert-Huang transform is applied to analyze single particle Lagrangian velocity data from
numerical simulations of hydrodynamic turbulence.   The velocity trajectory is described in terms of a set of intrinsic mode functions, $C_i(t)$, and of their instantaneous frequency, $\omega_i(t)$. 
On the basis of this decomposition we define the $\omega$-conditioned statistical moments of the $C_i$ 
modes, named $q$-order Hilbert Spectra (HS). 
We show that such new quantities have enhanced scaling properties as compared to 
traditional Fourier transform- or correlation-based (Structure Functions)
 statistical indicators, thus providing better 
insights into the turbulent energy transfer process. 
We  present a clear empirical evidence  that the energy-like quantity, i.e. the second-order HS, displays a linear scaling in time  in the inertial range, 
 as expected from dimensional analysis and never observed before. 
We also  measure high order moment scaling
exponents in a direct way, without resorting the Extended Self 
Similarity (ESS) procedure. This leads to a new estimate of the Lagrangian structure functions exponents which are consistent with the multifractal prediction in the 
Lagrangian frame as proposed in {\it [Biferale et al., Phys. Rev. Lett.  {\bf93}, 064502 
(2004)]}. 
  \end{abstract}

\maketitle
\end{CJK*}

The statistical description of a tracer trajectory in turbulent flows still lacks of a sound theoretical and phenomenological understanding \cite{Yeung2002ARFM,Toschi2009ARFM}.  Presently, no analytical results linking the Navier-Stokes  equations to the statistics of the velocity increments, $v(t+\tau)-v(t)$, along the particle evolution are known. On the ground of dimensional arguments, pure scaling laws are expected for time increments
larger than the Kolmogorov dissipative time, $\tau_\eta$, and smaller than the large-scale typical eddy-turn over time, $\mathcal{T}_L$. 
The ratio between the two time scales grows with the Reynolds number as $Re \propto \mathcal{T}_L/\tau_\eta$.  Despite of the many numerical 
and experimental  attempts \cite{grauer,YPS2006,Xu2006PRL,pinton,Chevillard2005}, 
no clear evidence of scaling properties have been detected in the Lagrangian domain even at high Reynolds numbers.
Such  a fact can be explained either invoking ultraviolet and infrared effects induced by the two cut-offs, $\tau_\eta$ and $ \mathcal{T}_L$ 
or by a real pure breaking of scaling invariance, independently of the Reynolds number \cite{Sawford2011PoF,Falkovich2012PoF}. 
Up to now, most of the attention has been payed to the so-called Lagrangian Structure Functions (LSF), i.e. moments of velocity
 increments:
\begin{equation}
S_q(\tau)=\langle \vert v_j(t+\tau)-v_j(t) \vert^q \rangle,
\end{equation}
where for simplicity we have assumed isotropy and dropped  any possible dependency of the l.h.s on the component of the velocity field. 
Phenomenological arguments based on a `bridge' relation between Eulerian and Lagrangian statistics \cite{Borgas,bof02,Biferale2004PRL,Schmitt2005,grauer_mhd,He2011PRE,biferale2011multi} 
predicts the existence of scaling properties also in the Lagrangian domain: 
$S_q(\tau) \sim \tau^{\zeta_L(q)}$ for $ \tau_\eta \ll \tau \ll \mathcal{T}_L$, with 
$\zeta_L(q)$ being related to the Eulerian scaling 
exponents, $\zeta_E(q)$, defining the scaling properties of  velocity increments between two points in
 the laboratory reference frame. Such relation has been well verified in the limit of very small time increments, by studying the statistics of flow acceleration \cite{Biferale2004PRL} or by using relative scaling properties \cite{Arneodo2008PRL}, i.e. studying one moment versus another one, a procedure known as ESS \cite{ESS}. On the other hand, no clear evidence of direct scaling properties as a function of $\tau$ has ever been 
detected (see \cite{Falkovich2012PoF,Sawford2011PoF} for two recent papers  discussing this problem). 
  \begin{figure}
\begin{center}
 \includegraphics[width=0.85\linewidth,clip]{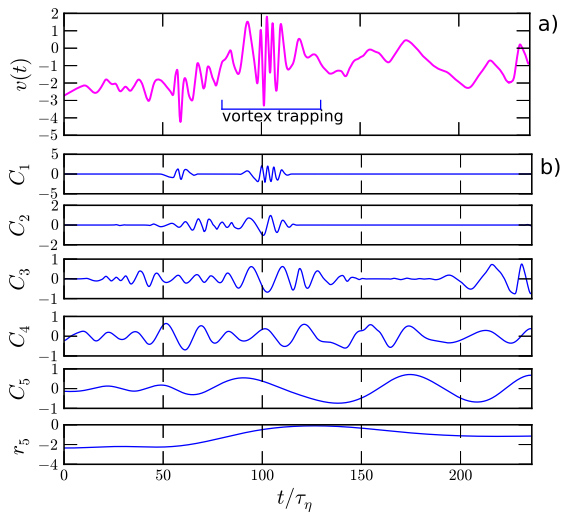}
   \vspace{-0.3cm}
  \caption{(a) An example of Lagrangian velocity $v(t)$ with vortex trapping event from  the DNS simulation. The data shows the
 multiscale nature of  Lagrangian turbulence with different time scales (structures) superimposed to each other. 
(b) Example of the decomposition of the above trajectory in intrinsic mode functions from empirical mode  decomposition. 
Note that the Lagrangian velocity is separated into different  functions with different time scales.  
The empirical mode decomposition approach reveals the multiscale property of the Lagrangian velocity  at a  local level.
}\label{fig:Trajectory}
\end{center}
\vspace{-1.0cm}
\end{figure}
As a result, despite the successful comparisons, using ESS, between theoretical predictions for $\zeta_{L}(q)/\zeta_{L}(2)$ and numerical and experimental Lagrangian measurements (see \cite{Arneodo2008PRL}), the absence of a clear scaling-range \textit{in the time domain} has cast doubts on the one side on the correctness and accuracy of the present phenomenological models, and on the other side
 on the fact that SF may not be the 
suitable statistical indicator to study turbulent flows in the Lagrangian domain 
\cite{Falkovich2012PoF}. One of the main concern regards possible non-local effects 
induced by either large-scales and low-frequencies modes or by small-scales and 
high-frequencies events that may result in sub-leading spurious contributions. It is 
well known for example that the temporal evolution of the velocity field  along a 
Lagrangian trajectory in a turbulent flows is strongly influenced by the presence of 
small-scales vortex filaments inducing visible high-frequency oscillations even on 
the single particle velocity signal (see Fig.\,\ref{fig:Trajectory} and 
Ref.\,\cite{Biferale2005PoF}).
In this paper we want to apply for the first time a relatively novel technique, called 
Hilbert-Huang Transform (HHT), to analyze multi-scale and multi-frequency signals 
which has revealed to be particularly useful in the data analysis of many complex 
systems \cite{Zhu1997,Loutridis2004,Cummings2004,Huang2005,Wu2007,Chen2010AADA}. 
HHT  has been recently applied  to analyze Eulerian 
turbulent data \cite{Huang2008EPL,Huang2010PRE,Huang2011PRE}, showing an 
unexpected ability to disentangle multiscale contributions. 
The main novelty of HHT relies on  its frequency-amplitude {\it adaptive} nature, being based on the decomposition of the original signal on a set of {\it quasi-eigenmodes} that are {\it not defined  a priori} \cite{Huang1998EMD,Flandrin2004EMDa}. The idea is to {\it not introduce} in the analysis any systematic pre-defined structures as it always happens using Fourier-based methodologies (e.g. Fourier decomposition or wavelet transforms).

In this paper, we apply and generalize the HHT methodology to extract the hierarchy of Lagrangian scaling exponent  $\zeta_{L}(q)$. The method is applied to the fluid trajectory data obtained from Direct Numerical Simulations (DNS) at $Re_{\lambda}=400$ (see Fig.\,\ref{fig:Trajectory}).  
We present a clear empirical evidence of scaling properties in the 
usual sense, as a power of the analyzed frequency, also in the Lagrangian domain.
 We show that the measured Hilbert-based  moments, $\mathcal{L}_q(\omega)$, display 
 a clear power law  on the range $0.01<\omega\tau_{\eta}<0.1$ at least up to the maximum order allowed to be measured by our statistics, $0\le q\le 4$. 
The exponents are in good quantitative agreement with the 
one predicted by using the `bridge relation' based on multifractal phenomenology \cite{Biferale2004PRL}, supporting even more the close relationship between Eulerian and Lagrangian fluctuations at least for what concerns velocity increments in 3D isotropic and homogeneous 
turbulence. 
The dataset considered here is composed by Lagrangian velocity trajectories in a   homogeneous and isotropic turbulent flow obtained from a $2048^3$ ($Re_{\lambda}=400$)  DNS simulation (more details in \cite{benzi2009velocity}). 
We analyze all the available $\sim 2\cdot10^5$
fluid tracer trajectories, each composed by  $N=4720$ time sampling of $v_j(t)  $ (where $j=1,2,3$ denotes the 
three velocity components) saved every
$0.1 \tau_{\eta}$ time units.
Therefore, we can access time scale from $0.1<\tau/\tau_{\eta}<236$, corresponding to the frequency range $0.004<\omega\tau_{\eta}<10$. 
  \begin{figure}
\begin{center}
 \includegraphics[width=0.85\linewidth,clip]{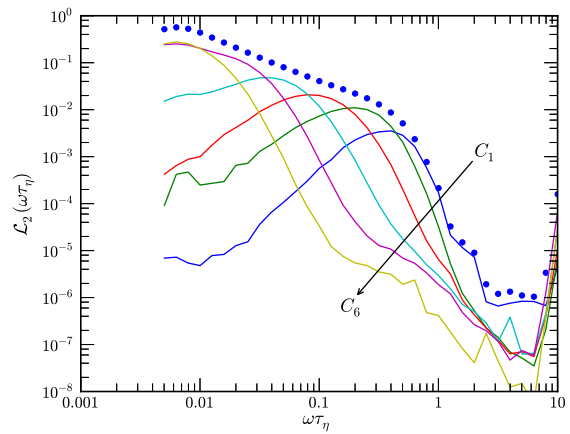}
  \vspace{-0.2cm}
  \caption{
Log-log plot of the second order Hilbert Spectrum, $\mathcal{L}_2(\omega) \equiv \Sigma_{i=1}^{n} \langle |C_i|^2 \vert\omega \rangle_{t}$, superposed with the different contributions from each IMF, $\langle (C_i)^2|\omega\rangle$ with $i=1,...,6$.}
\label{fig:L2}
\end{center}
\vspace{-1.0cm}
\end{figure}
The HHT is a procedure composed by two steps. 
The first step is the decomposition of the signal into its Intrinsic Mode Functions (IMF) followed by
 the Hilbert transform on such modes. In the first step, through a procedure called Empirical Mode Decomposition (EMD), we decompose each  velocity time series into the sum: 
 \vspace{-0.2cm}
\begin{equation}\label{decomp}
v(t)=\sum_{i=1}^{n} C_i(t)+r_n(t),
\end{equation} 
where $C_i(t)$ are the IMFs and $r_n(t)$ is a small residual, 
an almost constant function characterized by having at most one extreme along the whole trajectory (which will therefore be neglected in the following analysis) \cite{Huang1998EMD,Flandrin2004EMDa}. In eq. (\ref{decomp}) $n$ may depend on the trajectory, with a maximum value which is linked to its length as $n_{max}= \log_2(N) \simeq12$. Given the actual length of our trajectories,  with $n \simeq 6 -7$ 
we are typically able to reconstruct the full behaviors (see Fig.\,\ref{fig:Trajectory}).

To be an IMF, each $C_i(t)$  must satisfy the following two conditions: (1)
 the difference between the number of local extrema and the number 
 of zero-crossings must be zero or one; (2) the running mean value 
 of the envelope defined by the local maxima and the envelope 
 defined by the local minima is zero. 
Indeed, the IMF is an approximation of the so-called mono-component signal, which possesses a well 
 defined instantaneous frequency 
\cite{Huang1998EMD,Huang1999EMD}. 
The physical meaning of such 
decomposition is clear:  we want to 
decompose the original trajectory into {\it quasi-eigenmodes} with locally homogeneous 
oscillating properties
\cite{Huang1998EMD,Rilling2003EMD}. 
In the second step, one performs a 
Hilbert transform for each one of the IMFs, 
\begin{equation}
\overline{C}_i(t)=\frac{1}{\pi}P\int \frac{C_i(t')}{t-t'}\upd t', 
\end{equation}
where $P$ stands for the Cauchy principal value.
This allows to retrieve the {\it instantaneous} frequency associated to each $C_i$  via
 \vspace{-0.2cm}
\begin{equation}
\omega_i(t)={\frac{1}{2\pi}} \frac{\upd}{\upd t} \arctan \left(\frac{\overline{C}_i(t)}{C_i(t)}\right)
\end{equation} \cite{Huang1998EMD}.
Therefore, we construct the pair of functions $[C_i(t),\omega_i(t)]$ for all IMF 
modes, and this concludes the standard HHT procedure. Let us stress again the fully 
{\it adaptive} nature of the HHT, the IMFs are not defined {\it a priori}, and they 
accommodate the oscillatory degree of the analyzed signal without postulating 
systematic ``structures'' \cite{Huang1998EMD,Flandrin2004EMDa}. The most 
important consequence is that the HHT is typically free of sub-harmonics \cite{Huang2005,Huang2010PRE,Huang2011PRE}.
Here, in order to investigate the amplitude of turbulent velocity fluctuations versus their characteristic frequency, 
we define the $\omega$-dependent $q$-order statistical moment, $\mathcal{L}_q(\omega)$,  by computing the moments of each IMF conditioned on those instant of time where the corresponding {\it instantaneous} frequency has a given value, $\omega_i(t) =  \omega$  \begin{equation}
\mathcal{L}_q(\omega) \equiv \Sigma_{i=1}^{n} \langle \vert C_i\vert^q \vert\omega \rangle_{t},
\end{equation}
where $q\ge 0$ is a real number, and with $\langle \ldots \rangle_{t}$ we denote time- and ensemble-averaging
 over different trajectory realizations. We dub it Hilbert spectrum (HS) of order $q$. 
Let us notice that each HS can be seen as a superposition of
spectra obtained from different IMFs. 
\begin{figure}[!ht]
\begin{center}
\vspace{1.0cm}
 \includegraphics[width=0.85\linewidth,clip]{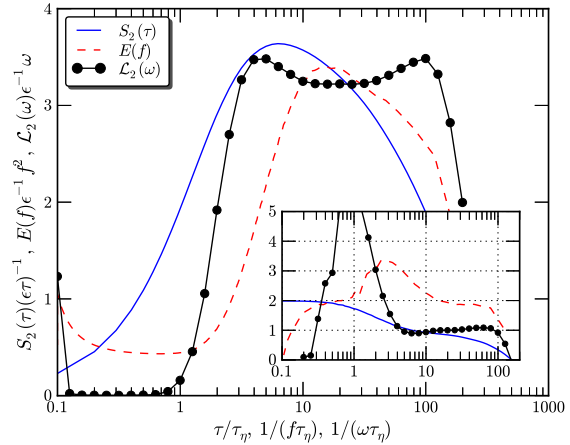}
 \caption{Comparison between the second-order compensated Lagrangian Structure Function $S_2(\tau) / (\epsilon \tau)\  \textit{vs}.\ \tau/\tau_{\eta}$ (solid line), the compensated Fourier spectrum $E(f)/\epsilon^{-1} f^2\  \textit{vs}.\ 1/(f \tau_{\eta})$ (dashed line) and the 
corresponding Hilbert spectrum $\mathcal{L}_2(\omega)\epsilon^{-1} \omega \ \textit{vs.}\ 1/(\omega \tau_{\eta})$ ($\bullet$), where $\tau_{\eta}$ represents the dissipative time scale of the turbulent flow and $\epsilon$ the mean energy dissipation rate.
In the below inset,  the logarithmic local slopes for $\upd \log{S_2(\tau)} / \upd \log{\tau}\ \textit{vs}.\ \tau/\tau_{\eta}$, 
$\upd \log{E(f)}/ \upd \log{f}\  \textit{vs}.\ 1/(f \tau_{\eta})$ and $\upd \log{\mathcal{L}_2(\omega)} / \upd \log{\omega}\ \textit{vs}.\ 1/(\omega\tau_{\eta})$. 
Note that the expected inertial scaling exponents are respectively:
$S_2(\tau) \sim \tau^{\zeta_L(2)}$, $E(f) \sim f^{-(\zeta_L(2)+1)}$ and $\mathcal{L}_2(\omega) \sim \omega^{-\zeta_L(2)}$,  with $\zeta_L(2)=1$.
} \label{fig:L2-S2}
 \end{center}
 \vspace{-1.0cm}
\end{figure}
%
\begin{figure}[!ht]
\begin{center}
\vspace{0.5cm}
 \includegraphics[width=0.85\linewidth,clip]{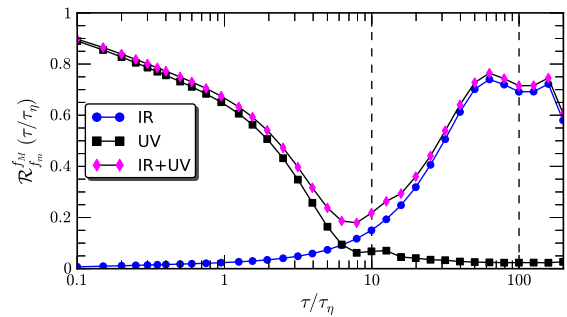}
 \caption{Relative contribution of Fourier frequencies  in the range $\left[ f_m, f_M \right]$ to the $S_2(\tau)$ LSF, as from eq.(\ref{eq:SFcontribution}). Low (IR) frequencies  $[0,10^{-2}]\tau_{\eta}^{-1}$  and high (UV) frequencies  $[10^{-1},+\infty]\tau_{\eta}^{-1}$. Vertical lines denote the empirically defined inertial range.
}\label{fig:IR-UV}
 \end{center}
 \vspace{-1.0cm}
\end{figure}

From a dimensional point of view the simplest link between the instantaneous frequency $\omega$ and the coherence time of an eddy $\tau$, is  the reciprocal relation $\omega \sim \tau^{-1}$. Therefore, we postulate for the general HS of order $q$ a scaling relation of the form 
\begin{equation}
\mathcal{L}_q(\omega)\sim \omega^{-\zeta_{L}(q)},
\label{eq:scaling}
\end{equation}
here,  $\zeta_{L}(q)$ must be compared with the scaling exponent provided by the LSF \cite{Huang2011PRE}.  We have validated the above scaling relation by using both fractional Brownian motion with various Hurst number $0<H<1$ for mono-fractal processes
 and a  lognormal signal with an intermittent parameter $\mu=0.15$ as an example of a multifractal process.
 For all cases, the scaling exponents provided by the HHT agree with the ones derived by standard SF method and with the theoretical ones \cite{Huang2011PRE}.
To begin with, we focus on the case $q=2$, that, as  mentioned, is related to the amplitude of  energy fluctuations as a function of its 
coherence time or characteristic frequency. In Fig.\,\ref{fig:L2} we show the second order HS, $\mathcal{L}_2(\omega)$ vs $\omega$ in log-log, superposed with the contributions from each different IMF order. As one can see, only the whole reconstructed HS shows a good scaling behaviour. 
  In order to better compare  HS to LSF curves we plot them in Fig.\,\ref{fig:L2-S2} in compensated form in such a  way that 
the expected behavior in the inertial range would be given by a  constant, respectively  $S_2(\tau)  (\epsilon \tau)^{-1}\  \textit{vs}.\ \tau$  and $\mathcal{L}_2(\omega)\epsilon^{-1}\omega \ \textit{vs.}\ 1/\omega$. For completeness in the same figure also the  compensated behavior of the Fourier spectrum, $E(f)\epsilon^{-1}f^2 \  \textit{vs}.\ 1/f$, is provided.
The first striking difference between HS and LSF or Fourier is the enhanced scaling property of the new quantity.
We also note that the shape of LSF curve is consistent  with the one in 
\cite{Falkovich2012PoF,Sawford2011PoF}, where no plateau was observed in the 
inertial range. On the compensated scale the Fourier spectrum behaves better than the LSF, 
but the range of scaling is about half of that of the Hilbert Spectrum. Such a 
difference is even more evident when the logarithmic local slopes are compared (see inset of Fig.\,\ref{fig:L2-S2}). 
A clear inertial scaling range, $0.01< \omega \tau_{\eta} <0.2$, corresponding to an interval of time scales $5 <\tau/\tau_{\eta}<100$, is observed for the compensated $\mathcal{L}_2$. 
The reason why LSF fails in displaying scaling, is that it mixes low (infrared, IR)/high (ultraviolet, UV)  frequency fluctuations to the ones in the inertial-range $\sim [10^{-2},10^{-1}]\tau_{\eta}^{-1}$. This becomes explicit when considering the relation, $S_2(\tau) \propto \int_{0}^{+\infty} E(f) (1 - \cos{2\pi f \tau})\upd f$, and defining
\begin{equation}\label{eq:SFcontribution}
\mathcal{R}_{f_{m}}^{f_{M}}(\tau) \equiv  S_2(\tau)^{-1} \int_{f_{m}}^{f_{M}} E(f') (1 - \cos{(2\pi f' \tau)}) \upd f' ,
\end{equation}
which measures the relative contributions to $S_2(\tau)$ from the frequency range $\left[f_{m}, f_{M}\right]$.
When such an interval is set to $[0,10^{-2}]\tau_{\eta}^{-1}$ we get the low frequency contributions, and with $[10^{-1},+\infty]\tau_{\eta}^{-1}$ the high ones. In figure Fig.\,\ref{fig:IR-UV}, we show that such spurious non-local contributions can be as high as $80\%$.
\begin{figure}[!htb]
\begin{center}
\vspace{0.5cm}
 \includegraphics[width=0.85\linewidth,clip]{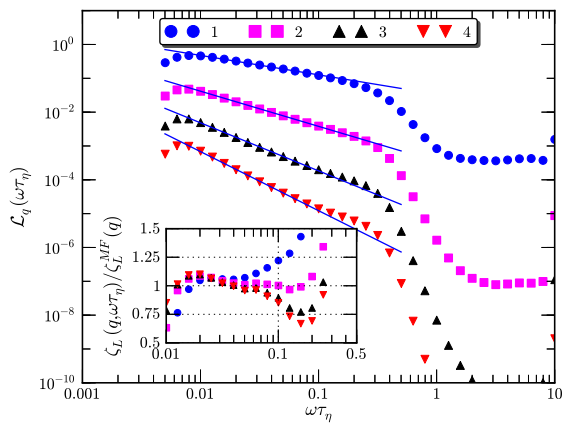}
   \vspace{-0.2cm}
 \caption{ 
The Hilbert spectra $\mathcal{L}_q(\omega\tau_{\eta})$ for $q=1,2,3,4$. For display clarity, the curves have been vertical shifted by factors $10^{-1}$, $10^{-2}$ and $10^{-3}$ for $q=2,\,3$ and 4. Solid lines comes from least square fit in the range is $0.01<\omega\tau_{\eta}<0.1$. The inset shows the comparison of the measured local scaling exponent $\zeta_L(q,\omega\tau_{\eta})= \upd \log{\mathcal{L}_q(\omega)}/\upd \log{\omega}$ with the multifractal  prediction $\zeta_{L}^{MF}(q)$.
}\label{fig:velocity}
\end{center}
\vspace{-0.0cm}   
\end{figure}
 \vspace{-0.6cm}
 
The HS functions $\mathcal{L}_q(\omega)$ have good scaling properties also for other $q$ orders. 
We calculated $\mathcal{L}_q(\omega)$ for the orders $q =1,2,3,4$,  and empirically found a good power law behavior
on the range $0.01<\omega\tau_{\eta}<0.1$ (resp. $10<\tau/\tau_{\eta}<100$), as shown in Fig.\,\ref{fig:velocity}.  
This allows to extract the scaling exponents directly in the instantaneous frequency space,
 without resorting to the above mentioned  ESS procedure. 
The numerical values for the $\zeta_L(q)$ extracted from the fit in the range  $0.01<\omega\tau_{\eta}<0.1$  are reported in the table \ref{tb:scaling}.   The values of the  scaling exponents are  estimated as the average of the logarithmic local slope $\zeta_L(q,\omega) = \upd \log{\mathcal{L}_2(\omega)} / \upd \log{\omega}$,  on the above interval
and the error bars as the difference between the averages taken on only the first or the second half (in log scale) of the fitted frequency range.
Note that the indicated errors are larger than the estimated statistical errors. Statistical convergence was here checked by performing the same analysis on random subsets with 1/64 of the total data.
First, let us notice the evident departure from the dimensional estimate (named K41 \cite{Monin1971}), $\zeta_L^{K41}(q)=q/2$.
Second, the measured values are in good agreement  with the  prediction given by the Multifractal model, $\zeta_{L}^{MF}$  \cite{Biferale2004PRL}.
In order to better appreciate the quality of our scaling, we show in the the inset of Fig.\,\ref{fig:velocity} 
the logarithmic local slope  empirically measured with the HHT, $\zeta_L(q,\omega)$, compensated with the predicted value
from the multifractal phenomenology, such that a plateau around the value $1$ is the indication of the existence of an 
 intermittent multifractal power law behavior. 

 \begin{table}[!htb]
 \begin{center}
 \begin{tabular}{| l|c| c| c| c| c| c| }
 \hline
    & $q=1$ & $q=2$ & $q=3$& $q=4$ \\
 \hline
  $\zeta_L^{K41}(q)$ &$0.5$ & $1.0$ & $1.5$ &$2.0$ \\
 \hline
 $\zeta_L^{MF}(q)$  &$0.55$ & $ 1 $ & $1.38$ &$1.71$ \\ 
 \hline
  $\zeta_L^{HS}(q)$ & $0.59 \pm0.06$&    $1.03 \pm0.03$ &   $1.39 \pm0.10$ &  $ 1.70\pm0.14$  \\
\hline
     \end{tabular}
 \caption{Lagrangian scaling exponents $\zeta_{L}(q)$ for orders $q=1,4$  as estimated from 
 dimensional analysis $q/2$ (K41), from the Multifractal model (MF) 
 \cite{Biferale2004PRL}, and as obtained here from Hilbert Spectra (HS). }
 \label{tb:scaling}
  \end{center}
  \vspace{-0.6cm}
 \end{table}


In summary,  we have presented a new Hilbert-Huang Transform based methodology to capture the intermittent nature of the turbulent Lagrangian
 velocity fluctuations. Our test bench has been a numerical database of homogeneous isotropic turbulence at $Re_{\lambda}=400$. 
The first remarkable result is that for the second-order statistical moment $\mathcal{L}_2(\omega)$, an energy-like quantity, we observe a clear inertial range versus time defined as $\tau=\omega^{-1}$ for at least one decade, in the  range $0.01<\omega\tau_{\eta}<0.2$. Such clean scaling has never been detected before using more standard methods.     
Second, we extracted the hierarchy of scaling exponent $\zeta_{L}(q)$ for the first time without applying ESS.  Our measurements provide a solid confirmation to the predictions of the multifractal model.
 The Hilbert method we propose in this paper is general and can be applied to other systems with multiscale dynamics, e.g., Rayleigh-B\'enard convection \cite{Lohse2010}, two dimensional turbulence \cite{Boffetta2012ARFM,Bouchet2012PhysRep}.
\begin{acknowledgments}
 This work is  sponsored  by the National Natural Science Foundation of China  (grant n$^{\circ}$ 11072139, 11032007 \& 11202122),`Pu Jiang' project of Shanghai (n$^{\circ}$ 12PJ1403500),  the Shanghai Program for Innovative Research Team in Universities,  COST Action MP0806  `Particles in turbulence' and in part by Foundation for Fundamental Research on Matter (FOM), which is part of the Netherlands Organization for Scientific Research (NWO). 
 The DNS data used in this study  are freely available from the
iCFDdatabase \footnote{http://cfd.cineca.it}.
 The EMD \textsc{Matlab} codes used in this study are
written by G. Rilling and P. Flandrin, Laboratoire Physique ENS
Lyon (France) \footnote{ http://perso.ens-lyon.fr/patrick.flandrin/emd.html}.
\end{acknowledgments}
\vspace{-0.5cm}


\begin{thebibliography}{39}%
\makeatletter
\providecommand \@ifxundefined [1]{%
 \@ifx{#1\undefined}
}%
\providecommand \@ifnum [1]{%
 \ifnum #1\expandafter \@firstoftwo
 \else \expandafter \@secondoftwo
 \fi
}%
\providecommand \@ifx [1]{%
 \ifx #1\expandafter \@firstoftwo
 \else \expandafter \@secondoftwo
 \fi
}%
\providecommand \natexlab [1]{#1}%
\providecommand \enquote  [1]{``#1''}%
\providecommand \bibnamefont  [1]{#1}%
\providecommand \bibfnamefont [1]{#1}%
\providecommand \citenamefont [1]{#1}%
\providecommand \href@noop [0]{\@secondoftwo}%
\providecommand \href [0]{\begingroup \@sanitize@url \@href}%
\providecommand \@href[1]{\@@startlink{#1}\@@href}%
\providecommand \@@href[1]{\endgroup#1\@@endlink}%
\providecommand \@sanitize@url [0]{\catcode `\\12\catcode `\$12\catcode
  `\&12\catcode `\#12\catcode `\^12\catcode `\_12\catcode `\%12\relax}%
\providecommand \@@startlink[1]{}%
\providecommand \@@endlink[0]{}%
\providecommand \url  [0]{\begingroup\@sanitize@url \@url }%
\providecommand \@url [1]{\endgroup\@href {#1}{\urlprefix }}%
\providecommand \urlprefix  [0]{URL }%
\providecommand \Eprint [0]{\href }%
\providecommand \doibase [0]{http://dx.doi.org/}%
\providecommand \selectlanguage [0]{\@gobble}%
\providecommand \bibinfo  [0]{\@secondoftwo}%
\providecommand \bibfield  [0]{\@secondoftwo}%
\providecommand \translation [1]{[#1]}%
\providecommand \BibitemOpen [0]{}%
\providecommand \bibitemStop [0]{}%
\providecommand \bibitemNoStop [0]{.\EOS\space}%
\providecommand \EOS [0]{\spacefactor3000\relax}%
\providecommand \BibitemShut  [1]{\csname bibitem#1\endcsname}%
\let\auto@bib@innerbib\@empty
\bibitem [{\citenamefont {Yeung}(2002)}]{Yeung2002ARFM}%
  \BibitemOpen
  \bibfield  {author} {\bibinfo {author} {\bibfnamefont {P.}~\bibnamefont
  {Yeung}},\ }\href@noop {} {\bibfield  {journal} {\bibinfo  {journal} {Annu.
  Rev. Fluid Mech.}\ }\textbf {\bibinfo {volume} {34}},\ \bibinfo {pages} {115}
  (\bibinfo {year} {2002})}\BibitemShut {NoStop}%
\bibitem [{\citenamefont {Toschi}\ and\ \citenamefont
  {Bodenschatz}(2009)}]{Toschi2009ARFM}%
  \BibitemOpen
  \bibfield  {author} {\bibinfo {author} {\bibfnamefont {F.}~\bibnamefont
  {Toschi}}\ and\ \bibinfo {author} {\bibfnamefont {E.}~\bibnamefont
  {Bodenschatz}},\ }\href@noop {} {\bibfield  {journal} {\bibinfo  {journal}
  {Annu. Rev. Fluid Mech.}\ }\textbf {\bibinfo {volume} {41}},\ \bibinfo
  {pages} {375} (\bibinfo {year} {2009})}\BibitemShut {NoStop}%
\bibitem [{\citenamefont {Kamps}\ \emph {et~al.}(2009)\citenamefont {Kamps},
  \citenamefont {Friedrich},\ and\ \citenamefont {Grauer}}]{grauer}%
  \BibitemOpen
  \bibfield  {author} {\bibinfo {author} {\bibfnamefont {O.}~\bibnamefont
  {Kamps}}, \bibinfo {author} {\bibfnamefont {R.}~\bibnamefont {Friedrich}}, \
  and\ \bibinfo {author} {\bibfnamefont {R.}~\bibnamefont {Grauer}},\
  }\href@noop {} {\bibfield  {journal} {\bibinfo  {journal} {Phys. Rev. E}\
  }\textbf {\bibinfo {volume} {79}},\ \bibinfo {pages} {066301} (\bibinfo
  {year} {2009})}\BibitemShut {NoStop}%
\bibitem [{\citenamefont {Yeung}\ \emph {et~al.}(2006)\citenamefont {Yeung},
  \citenamefont {Pope},\ and\ \citenamefont {Sawford}}]{YPS2006}%
  \BibitemOpen
  \bibfield  {author} {\bibinfo {author} {\bibfnamefont {P.~K.}\ \bibnamefont
  {Yeung}}, \bibinfo {author} {\bibfnamefont {S.~B.}\ \bibnamefont {Pope}}, \
  and\ \bibinfo {author} {\bibfnamefont {B.~L.}\ \bibnamefont {Sawford}},\
  }\href@noop {} {\bibfield  {journal} {\bibinfo  {journal} {J. Turb.}\
  }\textbf {\bibinfo {volume} {7}},\ \bibinfo {pages} {58} (\bibinfo {year}
  {2006})}\BibitemShut {NoStop}%
\bibitem [{\citenamefont {Xu}\ \emph {et~al.}(2006)\citenamefont {Xu},
  \citenamefont {Bourgoin}, \citenamefont {Ouellette},\ and\ \citenamefont
  {Bodenschatz}}]{Xu2006PRL}%
  \BibitemOpen
  \bibfield  {author} {\bibinfo {author} {\bibfnamefont {H.}~\bibnamefont
  {Xu}}, \bibinfo {author} {\bibfnamefont {M.}~\bibnamefont {Bourgoin}},
  \bibinfo {author} {\bibfnamefont {N.}~\bibnamefont {Ouellette}}, \ and\
  \bibinfo {author} {\bibfnamefont {E.}~\bibnamefont {Bodenschatz}},\
  }\href@noop {} {\bibfield  {journal} {\bibinfo  {journal} {Phys. Rev. Lett.}\
  }\textbf {\bibinfo {volume} {96}},\ \bibinfo {pages} {024503} (\bibinfo {year}
  {2006})}\BibitemShut {NoStop}%
\bibitem [{\citenamefont {Mordant}\ \emph {et~al.}(2003)\citenamefont
  {Mordant}, \citenamefont {Delour}, \citenamefont {L\'eveque}, \citenamefont
  {Michel}, \citenamefont {Arn\'eodo},\ and\ \citenamefont {Pinton}}]{pinton}%
  \BibitemOpen
  \bibfield  {author} {\bibinfo {author} {\bibfnamefont {N.}~\bibnamefont
  {Mordant}}, \bibinfo {author} {\bibfnamefont {J.}~\bibnamefont {Delour}},
  \bibinfo {author} {\bibfnamefont {E.}~\bibnamefont {L\'eveque}}, \bibinfo
  {author} {\bibfnamefont {O.}~\bibnamefont {Michel}}, \bibinfo {author}
  {\bibfnamefont {A.}~\bibnamefont {Arn\'eodo}}, \ and\ \bibinfo {author}
  {\bibfnamefont {J.-F.}\ \bibnamefont {Pinton}},\ }\href@noop {} {\bibfield
  {journal} {\bibinfo  {journal} {J. Stat. Phys.}\ }\textbf {\bibinfo {volume}
  {113}},\ \bibinfo {pages} {5/6} (\bibinfo {year} {2003})}\BibitemShut
  {NoStop}%
\bibitem [{\citenamefont {Chevillard}\ \emph {et~al.}(2005)\citenamefont
  {Chevillard}, \citenamefont {Roux}, \citenamefont {L{\'e}v{\^e}que},
  \citenamefont {Mordant}, \citenamefont {Pinton},\ and\ \citenamefont
  {Arn{\'e}odo}}]{Chevillard2005}%
  \BibitemOpen
  \bibfield  {author} {\bibinfo {author} {\bibfnamefont {L.}~\bibnamefont
  {Chevillard}}, \bibinfo {author} {\bibfnamefont {S.}~\bibnamefont {Roux}},
  \bibinfo {author} {\bibfnamefont {E.}~\bibnamefont {L{\'e}v{\^e}que}},
  \bibinfo {author} {\bibfnamefont {N.}~\bibnamefont {Mordant}}, \bibinfo
  {author} {\bibfnamefont {J.-F.}\ \bibnamefont {Pinton}}, \ and\ \bibinfo
  {author} {\bibfnamefont {A.}~\bibnamefont {Arn{\'e}odo}},\ }\href@noop {}
  {\bibfield  {journal} {\bibinfo  {journal} {Phys. Rev. Lett.}\ }\textbf
  {\bibinfo {volume} {95}},\ \bibinfo {pages} {064501} (\bibinfo {year}
  {2005})}\BibitemShut {NoStop}%
\bibitem [{\citenamefont {Sawford}\ and\ \citenamefont
  {Yeung}(2011)}]{Sawford2011PoF}%
  \BibitemOpen
  \bibfield  {author} {\bibinfo {author} {\bibfnamefont {B.}~\bibnamefont
  {Sawford}}\ and\ \bibinfo {author} {\bibfnamefont {P.}~\bibnamefont
  {Yeung}},\ }\href@noop {} {\bibfield  {journal} {\bibinfo  {journal} {Phys.
  Fluids}\ }\textbf {\bibinfo {volume} {23}},\ \bibinfo {pages} {091704}
  (\bibinfo {year} {2011})}\BibitemShut {NoStop}%
\bibitem [{\citenamefont {Falkovich~\textit{et al.}}(2012)}]{Falkovich2012PoF}%
  \BibitemOpen
  \bibfield  {author} {\bibinfo {author} {\bibfnamefont {G.}~\bibnamefont
  {Falkovich~\textit{et al.}}},\ }\href@noop {} {\bibfield  {journal} {\bibinfo
   {journal} {Phys. Fluids}\ }\textbf {\bibinfo {volume} {24}},\ \bibinfo
  {pages} {055102} (\bibinfo {year} {2012})}\BibitemShut {NoStop}%
\bibitem [{\citenamefont {Borgas}(1993)}]{Borgas}%
  \BibitemOpen
  \bibfield  {author} {\bibinfo {author} {\bibfnamefont {M.~S.}\ \bibnamefont
  {Borgas}},\ }\href@noop {} {\bibfield  {journal} {\bibinfo  {journal} {Phil.
  Trans. R. Soc. London A}\ }\textbf {\bibinfo {volume} {342}},\ \bibinfo
  {pages} {379} (\bibinfo {year} {1993})}\BibitemShut {NoStop}%
\bibitem [{\citenamefont {Boffetta}\ \emph {et~al.}(2002)\citenamefont
  {Boffetta}, \citenamefont {De~Lillo},\ and\ \citenamefont
  {Musacchio}}]{bof02}%
  \BibitemOpen
  \bibfield  {author} {\bibinfo {author} {\bibfnamefont {G.}~\bibnamefont
  {Boffetta}}, \bibinfo {author} {\bibfnamefont {F.}~\bibnamefont {De~Lillo}},
  \ and\ \bibinfo {author} {\bibfnamefont {S.}~\bibnamefont {Musacchio}},\
  }\href@noop {} {\bibfield  {journal} {\bibinfo  {journal} {Phys. Rev. E}\
  }\textbf {\bibinfo {volume} {66}},\ \bibinfo {pages} {066307} (\bibinfo
  {year} {2002})}\BibitemShut {NoStop}%
\bibitem [{\citenamefont {Biferale}\ \emph {et~al.}(2004)\citenamefont
  {Biferale}, \citenamefont {Boffetta}, \citenamefont {Celani}, \citenamefont
  {Devenish}, \citenamefont {Lanotte},\ and\ \citenamefont
  {Toschi}}]{Biferale2004PRL}%
  \BibitemOpen
  \bibfield  {author} {\bibinfo {author} {\bibfnamefont {L.}~\bibnamefont
  {Biferale}}, \bibinfo {author} {\bibfnamefont {G.}~\bibnamefont {Boffetta}},
  \bibinfo {author} {\bibfnamefont {A.}~\bibnamefont {Celani}}, \bibinfo
  {author} {\bibfnamefont {B.}~\bibnamefont {Devenish}}, \bibinfo {author}
  {\bibfnamefont {A.}~\bibnamefont {Lanotte}}, \ and\ \bibinfo {author}
  {\bibfnamefont {F.}~\bibnamefont {Toschi}},\ }\href@noop {} {\bibfield
  {journal} {\bibinfo  {journal} {Phys. Rev. Lett.}\ }\textbf {\bibinfo
  {volume} {93}},\ \bibinfo {pages} {064502} (\bibinfo {year}
  {2004})}\BibitemShut {NoStop}%
\bibitem [{\citenamefont {Schmitt}(2005)}]{Schmitt2005}%
  \BibitemOpen
  \bibfield  {author} {\bibinfo {author} {\bibfnamefont {F.}~\bibnamefont
  {Schmitt}},\ }\href@noop {} {\bibfield  {journal} {\bibinfo  {journal} {Eur.
  Phys. J. B}\ }\textbf {\bibinfo {volume} {48}},\ \bibinfo {pages} {129}
  (\bibinfo {year} {2005})}\BibitemShut {NoStop}%
\bibitem [{\citenamefont {Homann}\ \emph {et~al.}(2009)\citenamefont {Homann},
  \citenamefont {Kamps}, \citenamefont {Friedrich},\ and\ \citenamefont
  {Grauer}}]{grauer_mhd}%
  \BibitemOpen
  \bibfield  {author} {\bibinfo {author} {\bibfnamefont {H.}~\bibnamefont
  {Homann}}, \bibinfo {author} {\bibfnamefont {O.}~\bibnamefont {Kamps}},
  \bibinfo {author} {\bibfnamefont {R.}~\bibnamefont {Friedrich}}, \ and\
  \bibinfo {author} {\bibfnamefont {R.}~\bibnamefont {Grauer}},\ }\href@noop {}
  {\bibfield  {journal} {\bibinfo  {journal} {New J. Phys.}\ }\textbf {\bibinfo
  {volume} {11}},\ \bibinfo {pages} {073020} (\bibinfo {year}
  {2009})}\BibitemShut {NoStop}%
\bibitem [{\citenamefont {He}(2011)}]{He2011PRE}%
  \BibitemOpen
  \bibfield  {author} {\bibinfo {author} {\bibfnamefont {G.}~\bibnamefont
  {He}},\ }\href@noop {} {\bibfield  {journal} {\bibinfo  {journal} {Phys. Rev.
  E}\ }\textbf {\bibinfo {volume} {83}},\ \bibinfo {pages} {025301} (\bibinfo
  {year} {2011})}\BibitemShut {NoStop}%
\bibitem [{\citenamefont {Biferale}\ \emph {et~al.}(2011)\citenamefont
  {Biferale}, \citenamefont {Calzavarini},\ and\ \citenamefont
  {Toschi}}]{biferale2011multi}%
  \BibitemOpen
  \bibfield  {author} {\bibinfo {author} {\bibfnamefont {L.}~\bibnamefont
  {Biferale}}, \bibinfo {author} {\bibfnamefont {E.}~\bibnamefont
  {Calzavarini}}, \ and\ \bibinfo {author} {\bibfnamefont {F.}~\bibnamefont
  {Toschi}},\ }\href@noop {} {\bibfield  {journal} {\bibinfo  {journal} {Phys.
  Fluids}\ }\textbf {\bibinfo {volume} {23}},\ \bibinfo {pages} {085107}
  (\bibinfo {year} {2011})}\BibitemShut {NoStop}%
\bibitem [{\citenamefont {Arn{\'e}odo~\textit{et al.}}(2008)}]{Arneodo2008PRL}%
  \BibitemOpen
  \bibfield  {author} {\bibinfo {author} {\bibfnamefont {A.}~\bibnamefont
  {Arn{\'e}odo~\textit{et al.}}},\ }\href@noop {} {\bibfield  {journal}
  {\bibinfo  {journal} {Phys. Rev. Lett.}\ }\textbf {\bibinfo {volume} {100}},\
  \bibinfo {pages} {254504} (\bibinfo {year} {2008})}\BibitemShut {NoStop}%
\bibitem [{\citenamefont {Benzi}\ \emph {et~al.}(1993)\citenamefont {Benzi},
  \citenamefont {Ciliberto}, \citenamefont {Tripiccione}, \citenamefont
  {Baudet}, \citenamefont {Massaioli},\ and\ \citenamefont {Succi}}]{ESS}%
  \BibitemOpen
  \bibfield  {author} {\bibinfo {author} {\bibfnamefont {R.}~\bibnamefont
  {Benzi}}, \bibinfo {author} {\bibfnamefont {S.}~\bibnamefont {Ciliberto}},
  \bibinfo {author} {\bibfnamefont {R.}~\bibnamefont {Tripiccione}}, \bibinfo
  {author} {\bibfnamefont {C.}~\bibnamefont {Baudet}}, \bibinfo {author}
  {\bibfnamefont {F.}~\bibnamefont {Massaioli}}, \ and\ \bibinfo {author}
  {\bibfnamefont {S.}~\bibnamefont {Succi}},\ }\href@noop {} {\bibfield
  {journal} {\bibinfo  {journal} {Phys. Rev. E}\ }\textbf {\bibinfo {volume}
  {48}},\ \bibinfo {pages} {29} (\bibinfo {year} {1993})}\BibitemShut {NoStop}%
\bibitem [{\citenamefont {Biferale}\ \emph {et~al.}(2005)\citenamefont
  {Biferale}, \citenamefont {Boffetta}, \citenamefont {Celani}, \citenamefont
  {Devenish}, \citenamefont {Lanotte},\ and\ \citenamefont
  {Toschi}}]{Biferale2005PoF}%
  \BibitemOpen
  \bibfield  {author} {\bibinfo {author} {\bibfnamefont {L.}~\bibnamefont
  {Biferale}}, \bibinfo {author} {\bibfnamefont {G.}~\bibnamefont {Boffetta}},
  \bibinfo {author} {\bibfnamefont {A.}~\bibnamefont {Celani}}, \bibinfo
  {author} {\bibfnamefont {B.}~\bibnamefont {Devenish}}, \bibinfo {author}
  {\bibfnamefont {A.}~\bibnamefont {Lanotte}}, \ and\ \bibinfo {author}
  {\bibfnamefont {F.}~\bibnamefont {Toschi}},\ }\href@noop {} {\bibfield
  {journal} {\bibinfo  {journal} {Phys. Fluids}\ }\textbf {\bibinfo {volume}
  {17}},\ \bibinfo {pages} {115101} (\bibinfo {year} {2005})}\BibitemShut
  {NoStop}%
\bibitem [{\citenamefont {Zhu}\ \emph {et~al.}(1997)\citenamefont {Zhu},
  \citenamefont {Shen}, \citenamefont {Eckermann}, \citenamefont {Bittner},
  \citenamefont {Hirota},\ and\ \citenamefont {Yee}}]{Zhu1997}%
  \BibitemOpen
  \bibfield  {author} {\bibinfo {author} {\bibfnamefont {X.}~\bibnamefont
  {Zhu}}, \bibinfo {author} {\bibfnamefont {Z.}~\bibnamefont {Shen}}, \bibinfo
  {author} {\bibfnamefont {S.~D.}\ \bibnamefont {Eckermann}}, \bibinfo {author}
  {\bibfnamefont {M.}~\bibnamefont {Bittner}}, \bibinfo {author} {\bibfnamefont
  {I.}~\bibnamefont {Hirota}}, \ and\ \bibinfo {author} {\bibfnamefont {J.~H.}\
  \bibnamefont {Yee}},\ }\href@noop {} {\bibfield  {journal} {\bibinfo
  {journal} {J. Geophys. Res}\ }\textbf {\bibinfo {volume} {102}},\ \bibinfo
  {pages} {16545} (\bibinfo {year} {1997})}\BibitemShut {NoStop}%
\bibitem [{\citenamefont {Loutridis}(2004)}]{Loutridis2004}%
  \BibitemOpen
  \bibfield  {author} {\bibinfo {author} {\bibfnamefont {S.~J.}\ \bibnamefont
  {Loutridis}},\ }\href@noop {} {\bibfield  {journal} {\bibinfo  {journal}
  {Eng. Struct.}\ }\textbf {\bibinfo {volume} {26}},\ \bibinfo {pages} {1833}
  (\bibinfo {year} {2004})}\BibitemShut {NoStop}%
\bibitem [{\citenamefont {Cummings}\ \emph {et~al.}(2004)\citenamefont
  {Cummings}, \citenamefont {Irizarry}, \citenamefont {Huang}, \citenamefont
  {Endy}, \citenamefont {Nisalak},\ and\ \citenamefont
  {Ungchusak}}]{Cummings2004}%
  \BibitemOpen
  \bibfield  {author} {\bibinfo {author} {\bibfnamefont {D.~A.~T.}\
  \bibnamefont {Cummings}}, \bibinfo {author} {\bibfnamefont {R.~A.}\
  \bibnamefont {Irizarry}}, \bibinfo {author} {\bibfnamefont {N.~E.}\
  \bibnamefont {Huang}}, \bibinfo {author} {\bibfnamefont {T.~P.}\ \bibnamefont
  {Endy}}, \bibinfo {author} {\bibfnamefont {A.}~\bibnamefont {Nisalak}}, \
  and\ \bibinfo {author} {\bibfnamefont {K.}~\bibnamefont {Ungchusak}},\
  }\href@noop {} {\bibfield  {journal} {\bibinfo  {journal} {Nature}\ }\textbf
  {\bibinfo {volume} {427}},\ \bibinfo {pages} {344} (\bibinfo {year}
  {2004})}\BibitemShut {NoStop}%
\bibitem [{\citenamefont {Huang}\ and\ \citenamefont {Wu}(2005)}]{Huang2005}%
  \BibitemOpen
  \bibfield  {author} {\bibinfo {author} {\bibfnamefont {N.~E.}\ \bibnamefont
  {Huang}}\ and\ \bibinfo {author} {\bibfnamefont {Z.~H.}\ \bibnamefont {Wu}},\
  }\href@noop {} {\bibfield  {journal} {\bibinfo  {journal} {Proceedings of the
  4th International Conference on Wavelet and Its Application, Macao}\ }
  (\bibinfo {year} {2005})}\BibitemShut {NoStop}%
\bibitem [{\citenamefont {Wu}\ \emph {et~al.}(2007)\citenamefont {Wu},
  \citenamefont {Huang}, \citenamefont {Long},\ and\ \citenamefont
  {Peng}}]{Wu2007}%
  \BibitemOpen
  \bibfield  {author} {\bibinfo {author} {\bibfnamefont {Z.~H.}\ \bibnamefont
  {Wu}}, \bibinfo {author} {\bibfnamefont {N.~E.}\ \bibnamefont {Huang}},
  \bibinfo {author} {\bibfnamefont {S.~R.}\ \bibnamefont {Long}}, \ and\
  \bibinfo {author} {\bibfnamefont {C.~K.}\ \bibnamefont {Peng}},\ }\href@noop
  {} {\bibfield  {journal} {\bibinfo  {journal} {PNAS}\ }\textbf {\bibinfo
  {volume} {104}},\ \bibinfo {pages} {14889} (\bibinfo {year}
  {2007})}\BibitemShut {NoStop}%
\bibitem [{\citenamefont {Chen}\ \emph {et~al.}(2010)\citenamefont {Chen},
  \citenamefont {Wu},\ and\ \citenamefont {Huang}}]{Chen2010AADA}%
  \BibitemOpen
  \bibfield  {author} {\bibinfo {author} {\bibfnamefont {X.~Y.}\ \bibnamefont
  {Chen}}, \bibinfo {author} {\bibfnamefont {Z.~H.}\ \bibnamefont {Wu}}, \ and\
  \bibinfo {author} {\bibfnamefont {N.~E.}\ \bibnamefont {Huang}},\ }\href@noop
  {} {\bibfield  {journal} {\bibinfo  {journal} {Adv. Adapt. Data Anal}\
  }\textbf {\bibinfo {volume} {2}},\ \bibinfo {pages} {233} (\bibinfo {year}
  {2010})}\BibitemShut {NoStop}%
\bibitem [{\citenamefont {Huang}\ \emph {et~al.}(2008)\citenamefont {Huang},
  \citenamefont {Schmitt}, \citenamefont {Lu},\ and\ \citenamefont
  {Liu}}]{Huang2008EPL}%
  \BibitemOpen
  \bibfield  {author} {\bibinfo {author} {\bibfnamefont {Y.}~\bibnamefont
  {Huang}}, \bibinfo {author} {\bibfnamefont {F.}~\bibnamefont {Schmitt}},
  \bibinfo {author} {\bibfnamefont {Z.}~\bibnamefont {Lu}}, \ and\ \bibinfo
  {author} {\bibfnamefont {Y.}~\bibnamefont {Liu}},\ }\href@noop {} {\bibfield
  {journal} {\bibinfo  {journal} {Europhys. Lett.}\ }\textbf {\bibinfo {volume}
  {84}},\ \bibinfo {pages} {40010} (\bibinfo {year} {2008})}\BibitemShut
  {NoStop}%
\bibitem [{\citenamefont {Huang}\ \emph {et~al.}(2010)\citenamefont {Huang},
  \citenamefont {Schmitt}, \citenamefont {Lu}, \citenamefont {Fougairolles},
  \citenamefont {Gagne},\ and\ \citenamefont {Liu}}]{Huang2010PRE}%
  \BibitemOpen
  \bibfield  {author} {\bibinfo {author} {\bibfnamefont {Y.}~\bibnamefont
  {Huang}}, \bibinfo {author} {\bibfnamefont {F.}~\bibnamefont {Schmitt}},
  \bibinfo {author} {\bibfnamefont {Z.}~\bibnamefont {Lu}}, \bibinfo {author}
  {\bibfnamefont {P.}~\bibnamefont {Fougairolles}}, \bibinfo {author}
  {\bibfnamefont {Y.}~\bibnamefont {Gagne}}, \ and\ \bibinfo {author}
  {\bibfnamefont {Y.}~\bibnamefont {Liu}},\ }\href@noop {} {\bibfield
  {journal} {\bibinfo  {journal} {Phys. Rev. E}\ }\textbf {\bibinfo {volume}
  {82}},\ \bibinfo {pages} {026319} (\bibinfo {year} {2010})}\BibitemShut
  {NoStop}%
\bibitem [{\citenamefont {Huang}\ \emph {et~al.}(2011)\citenamefont {Huang},
  \citenamefont {Schmitt}, \citenamefont {Hermand}, \citenamefont {Gagne},
  \citenamefont {Lu},\ and\ \citenamefont {Liu}}]{Huang2011PRE}%
  \BibitemOpen
  \bibfield  {author} {\bibinfo {author} {\bibfnamefont {Y.}~\bibnamefont
  {Huang}}, \bibinfo {author} {\bibfnamefont {F.~G.}\ \bibnamefont {Schmitt}},
  \bibinfo {author} {\bibfnamefont {J.-P.}\ \bibnamefont {Hermand}}, \bibinfo
  {author} {\bibfnamefont {Y.}~\bibnamefont {Gagne}}, \bibinfo {author}
  {\bibfnamefont {Z.}~\bibnamefont {Lu}}, \ and\ \bibinfo {author}
  {\bibfnamefont {Y.}~\bibnamefont {Liu}},\ }\href@noop {} {\bibfield
  {journal} {\bibinfo  {journal} {Phys. Rev. E}\ }\textbf {\bibinfo {volume}
  {84}},\ \bibinfo {pages} {016208} (\bibinfo {year} {2011})}\BibitemShut
  {NoStop}%
\bibitem [{\citenamefont {Huang}\ \emph {et~al.}(1998)\citenamefont {Huang},
  \citenamefont {Shen}, \citenamefont {Long}, \citenamefont {Wu}, \citenamefont
  {Shih}, \citenamefont {Zheng}, \citenamefont {Yen}, \citenamefont {Tung},\
  and\ \citenamefont {Liu}}]{Huang1998EMD}%
  \BibitemOpen
  \bibfield  {author} {\bibinfo {author} {\bibfnamefont {N.~E.}\ \bibnamefont
  {Huang}}, \bibinfo {author} {\bibfnamefont {Z.}~\bibnamefont {Shen}},
  \bibinfo {author} {\bibfnamefont {S.~R.}\ \bibnamefont {Long}}, \bibinfo
  {author} {\bibfnamefont {M.~C.}\ \bibnamefont {Wu}}, \bibinfo {author}
  {\bibfnamefont {H.~H.}\ \bibnamefont {Shih}}, \bibinfo {author}
  {\bibfnamefont {Q.}~\bibnamefont {Zheng}}, \bibinfo {author} {\bibfnamefont
  {N.}~\bibnamefont {Yen}}, \bibinfo {author} {\bibfnamefont {C.~C.}\
  \bibnamefont {Tung}}, \ and\ \bibinfo {author} {\bibfnamefont {H.~H.}\
  \bibnamefont {Liu}},\ }\href@noop {} {\bibfield  {journal} {\bibinfo
  {journal} {Proc. R. Soc. London, Ser. A}\ }\textbf {\bibinfo {volume}
  {454}},\ \bibinfo {pages} {903} (\bibinfo {year} {1998})}\BibitemShut
  {NoStop}%
\bibitem [{\citenamefont {Flandrin}\ and\ \citenamefont
  {Gon{\c{c}}alv\`es}(2004)}]{Flandrin2004EMDa}%
  \BibitemOpen
  \bibfield  {author} {\bibinfo {author} {\bibfnamefont {P.}~\bibnamefont
  {Flandrin}}\ and\ \bibinfo {author} {\bibfnamefont {P.}~\bibnamefont
  {Gon{\c{c}}alv\`es}},\ }\href@noop {} {\bibfield  {journal} {\bibinfo
  {journal} {Int. J. Wavelets, Multires. Info. Proc.}\ }\textbf {\bibinfo
  {volume} {2}},\ \bibinfo {pages} {477} (\bibinfo {year} {2004})}\BibitemShut
  {NoStop}%
\bibitem [{\citenamefont {Benzi}\ \emph {et~al.}(2009)\citenamefont {Benzi},
  \citenamefont {Biferale}, \citenamefont {Calzavarini}, \citenamefont
  {Lohse},\ and\ \citenamefont {Toschi}}]{benzi2009velocity}%
  \BibitemOpen
  \bibfield  {author} {\bibinfo {author} {\bibfnamefont {R.}~\bibnamefont
  {Benzi}}, \bibinfo {author} {\bibfnamefont {L.}~\bibnamefont {Biferale}},
  \bibinfo {author} {\bibfnamefont {E.}~\bibnamefont {Calzavarini}}, \bibinfo
  {author} {\bibfnamefont {D.}~\bibnamefont {Lohse}}, \ and\ \bibinfo {author}
  {\bibfnamefont {F.}~\bibnamefont {Toschi}},\ }\href@noop {} {\bibfield
  {journal} {\bibinfo  {journal} {Phys. Rev. E}\ }\textbf {\bibinfo {volume}
  {80}},\ \bibinfo {pages} {066318} (\bibinfo {year} {2009})}\BibitemShut
  {NoStop}%
\bibitem [{\citenamefont {Huang}\ \emph {et~al.}(1999)\citenamefont {Huang},
  \citenamefont {Shen},\ and\ \citenamefont {Long}}]{Huang1999EMD}%
  \BibitemOpen
  \bibfield  {author} {\bibinfo {author} {\bibfnamefont {N.~E.}\ \bibnamefont
  {Huang}}, \bibinfo {author} {\bibfnamefont {Z.}~\bibnamefont {Shen}}, \ and\
  \bibinfo {author} {\bibfnamefont {S.~R.}\ \bibnamefont {Long}},\ }\href@noop
  {} {\bibfield  {journal} {\bibinfo  {journal} {Annu. Rev. Fluid Mech.}\
  }\textbf {\bibinfo {volume} {31}},\ \bibinfo {pages} {417} (\bibinfo {year}
  {1999})}\BibitemShut {NoStop}%
\bibitem [{\citenamefont {Rilling}\ \emph {et~al.}(2003)\citenamefont
  {Rilling}, \citenamefont {Flandrin},\ and\ \citenamefont
  {Gon\c{c}alv\`es}}]{Rilling2003EMD}%
  \BibitemOpen
  \bibfield  {author} {\bibinfo {author} {\bibfnamefont {G.}~\bibnamefont
  {Rilling}}, \bibinfo {author} {\bibfnamefont {P.}~\bibnamefont {Flandrin}}, \
  and\ \bibinfo {author} {\bibfnamefont {P.}~\bibnamefont {Gon\c{c}alv\`es}},\
  }\href@noop {} {\bibfield  {journal} {\bibinfo  {journal} {IEEE-EURASIP
  Workshop on Nonlinear Signal and Image Processing}\ } (\bibinfo {year}
  {2003})}\BibitemShut {NoStop}%
\bibitem [{\citenamefont {Monin}\ and\ \citenamefont
  {Yaglom}(1971)}]{Monin1971}%
  \BibitemOpen
  \bibfield  {author} {\bibinfo {author} {\bibfnamefont {A.~S.}\ \bibnamefont
  {Monin}}\ and\ \bibinfo {author} {\bibfnamefont {A.~M.}\ \bibnamefont
  {Yaglom}},\ }\href@noop {} {\emph {\bibinfo {title} {{Statistical fluid
  mechanics vd II}}}}\ (\bibinfo  {publisher} {MIT Press Cambridge, Mass},\
  \bibinfo {year} {1971})\BibitemShut {NoStop}%
\bibitem [{\citenamefont {Lohse}\ and\ \citenamefont {Xia}(2010)}]{Lohse2010}%
  \BibitemOpen
  \bibfield  {author} {\bibinfo {author} {\bibfnamefont {D.}~\bibnamefont
  {Lohse}}\ and\ \bibinfo {author} {\bibfnamefont {K.-Q.}\ \bibnamefont
  {Xia}},\ }\href@noop {} {\bibfield  {journal} {\bibinfo  {journal} {Annu.
  Rev. Fluid Mech.}\ }\textbf {\bibinfo {volume} {42}},\ \bibinfo {pages} {335}
  (\bibinfo {year} {2010})}\BibitemShut {NoStop}%
\bibitem [{\citenamefont {Boffetta}\ and\ \citenamefont
  {Ecke}(2012)}]{Boffetta2012ARFM}%
  \BibitemOpen
  \bibfield  {author} {\bibinfo {author} {\bibfnamefont {G.}~\bibnamefont
  {Boffetta}}\ and\ \bibinfo {author} {\bibfnamefont {R.}~\bibnamefont
  {Ecke}},\ }\href@noop {} {\bibfield  {journal} {\bibinfo  {journal} {Annu.
  Rev. Fluid Mech}\ }\textbf {\bibinfo {volume} {44}},\ \bibinfo {pages} {427}
  (\bibinfo {year} {2012})}\BibitemShut {NoStop}%
\bibitem [{\citenamefont {Bouchet}\ and\ \citenamefont
  {Venaille}(2012)}]{Bouchet2012PhysRep}%
  \BibitemOpen
  \bibfield  {author} {\bibinfo {author} {\bibfnamefont {F.}~\bibnamefont
  {Bouchet}}\ and\ \bibinfo {author} {\bibfnamefont {A.}~\bibnamefont
  {Venaille}},\ }\href@noop {} {\bibfield  {journal} {\bibinfo  {journal}
  {Phys. Rep.}\ }\textbf {\bibinfo {volume} {515}},\ \bibinfo {pages} {227}
  (\bibinfo {year} {2012})}\BibitemShut {NoStop}%
\bibitem [{Note1()}]{Note1}%
  \BibitemOpen
  \bibinfo {note} {Http://cfd.cineca.it}\BibitemShut {NoStop}%
\bibitem [{Note2()}]{Note2}%
  \BibitemOpen
  \bibinfo {note}
  {Http://perso.ens-lyon.fr/patrick.flandrin/emd.html}\BibitemShut {NoStop}%
\end{thebibliography}
%
\end{document}